\documentclass[lnbip,sechang,a4paper]{svmultln}
\usepackage{amssymb}
\usepackage{graphicx}

\usepackage{url}
\urldef{\mailsa}\path|{Isabelle.Mirbel, Pierre.Crescenzo}@unice.fr|
\usepackage[pdfpagelabels,hypertexnames=false,breaklinks=true,bookmarksopen=true,bookmarksopenlevel=2]{hyperref}

\begin{document}

\mainmatter  

\title{From End-User's Requirements to Web Services Retrieval: a Semantic and Intention-Driven Approach}

\titlerunning{From End-User's Requirements to Web Services Retrieval}

%
%
\author{Isabelle Mirbel \and Pierre Crescenzo}  %
\authorrunning{I. Mirbel and P. Crescenzo}

\institute{Universit\'e de Nice Sophia-Antipolis\\
Laboratoire I3S (UNS/CNRS)\\
930 route des Colles\\
BP 145\\
F-06903 Sophia-Antipolis cedex\\
France\\
\mailsa\\
}

%
%

\toctitle{ }
\tocauthor{ }
\maketitle

\begin{abstract} 

In this paper, we present SATIS, a framework to derive Web Service
specifications from end-user's requirements in order to operationalise
business processes in the context of a specific application
domain. The aim of SATIS is to provide to neuroscientists, which are
not familiar with computer science, a complete solution to easily find
a set of Web Services to implement an image processing pipeline.  More
precisely, our framework offers the capability to capture high-level
end-user's requirements in an iterative and incremental way and to turn
them into queries to retrieve Web Services description. The whole
framework relies on reusable and combinable elements which can be
shared out by a community of users sharing some interest or problems
for a given topic.  In our approach, we adopt Web semantic
languages and models as a unified framework to deal with end-user's
requirements and Web Service descriptions in order to take advantage
of their reasoning and traceability capabilities.
 
 \end{abstract}

\section{Introduction}

Service-oriented computing is a paradigm relying on services
as atomic constructs to support the development and easy composition
of distributed applications. Application components are assembled with
little efforts into workflows of services loosely coupled to
operationalise flexible and dynamic business processes. Searching for
the relevant Web Services to operationalise a particular business
process is one of the challenges of the service-oriented computing
area. At present, in the process of searching for Web Services, it is
assumed that user's goals have already been identified, captured,
specified and formalised in a suitable model to easily find the
relevant services. Or it is considered that users, which often are
specialists of their domain, are also computer scientists or at least
connoisseurs of Web Services. These hypotheses are generally too
strong to be reasonable.

In this context, the SATIS ({\textit{Semantically AnnotaTed Intentions for
Services}) project's ambition is to allow final users
to express their intentions (or goals) and strategies (to achieve
their intentions) in a high-level language, and to support the
selection of a set of Web Service descriptions which could respond to
the users' needs. But this problem is complex and cannot be solved in
a general approach. Therefore, we focus on an application domain where
domain knowledge and service descriptions (semantic Web Services) are
available. The aim of SATIS is to provide to neuroscientists, which are not familiar with
computer science, a complete solution to easily find a set of Web
Services to implement an image processing pipeline.

Indeed, our purpose is to give at users disposal some useful dedicated
reusable fragments of know-how to help them to implement their business
goals with Web Services. Therefore, our approach relies on high-level
business-oriented activity specification with the help of an
intentional model in order to derive Web Service description from
this high-level specification. We also focus on a community of users
sharing some interest or problems for a given topic inside the
business domain.

Our work belongs to the family of goal-based service retrieval
approaches. These approaches
(\cite{Stollberg07,Vukovic05,Zhang06,Bonino08}) aim at specifying the
goals which have to be satisfied by the retrieved services. In these
proposals, different models are provided to specify goals without
addressing the problem of how to capture them. On the contrary, our
aim is to provide means to assist final users in querying the Web
Service registry to find Web Services to operationalise a business
process. The GODO approach \cite{Gomez06} also addresses this issue by
proposing models and tools to capture user's goals with the help of an
ontology or in natural language. As in \cite{Kaabi07}, we propose an
incremental process to refine users' requirements in order to specify
the features required for the Web Services under retrieval. Our
approach distinguishes itself from \cite{Kaabi07} by the fact that we
rely on semantic Web models and techniques to enrich the goal (or
intention) specification, in order to provide reasoning and
explanation capabilities.

With regards to approaches dealing with ontology-based service
discovery \cite{OWLS-07}, and more precisely OWL-S based approaches (as
we are relying on OWL-S with regards to Web Service descriptions),
capability matching algorithms \cite{41-OWLS} exploiting service profile
descriptions have been proposed. Matchmaking algorithms \cite{64-OWLS}
comparing state transformations described in the query to the ones
provided in the descriptions have also been proposed. All these
algorithms mainly exploit features of subsumption
relationships. Ranking mechanisms have also been provided \cite{8-OWLS}. Our
approach distinguishes itself from these works by the fact that our focus is on providing means to assist final users in authoring
queries (more than rendering them). In other words, we are interested
in the upstream process of deriving queries from final users
requirements. Moreover, our concern is also on how to annotate such
queries in order to support their capitalisation and sharing among a
community of users.

Beyond an
alternative way to search for Web Services, we provide
means to capitalise know-how about Web Service search
procedures themselves. Another novelty of our approach is to
operationalise goals by rules in order to promote both mutualisation
of high-level intentional specification and cross-fertilisation of
know-how about Web Services search procedures among the
community members.

The paper is organised as follows. First we give an overview of our
SATIS approach in section \ref{satis}. Then, in section
\ref{authoring}, we detail the authoring process proposed in SATIS and
how the authored search procedure is rendered in section
\ref{rendering}. Next, we explain in section \ref{jungle} how the
framework is used by the different actors interacting in a
neurosciences community of users. Finally, we conclude and give some
perspectives.

\section{SATIS approach}
\label{satis}

The aim of our approach is to provide to neuroscientists, which are not
familiar with computer science, a complete solution to easily find a
set of Web Services to implement an image processing pipeline. The
focus of this proposal is on how to search and retrieve Web Services
descriptions from end-user's requirements. Indeed we provide support to
retrieve an organised set of Web Service descriptions suitable to
operationalise an image processing pipeline as specified by a
neuroscientist (final user).

As we are interested by high-level end-user's requirements, we rely on a
dedicated graphical notation to capture and specify them. In the
context of a neuroscientists community, these requirements deal with
image analysis pipelines. Different business process modeling
formalisms have been proposed in the
literature \cite{Nurcan05}. Decision-oriented models are semantically
more powerful than the other process models because they explain not
only how the process proceeds but also why. Their enactment guides the
decision making process that shapes the process, and helps reasoning
about the rationale \cite{Nurcan05}. Our approach is based on the
adaptation of such a decision-oriented model called the map
model \cite{Rolland07}. This intentional process modeling formalism
allows final users (neuroscientists) to define their image analysis
pipeline by describing intermediate \emph{intentions} (\textit{i.e.} goals and
subgoals to be satisfied through the processing chain) and
\emph{strategies} (\textit{i.e.} means to reach goals).

As we are interested by the end-user's point of view on the processing
pipeline to be operationalised by Web Services, we don't want him/her
to explicitly specify the Web Service(s) s/he is interested in but the
intention(s) s/he wants to satisfy by rendering Web Service(s). Moreover,
we don't want to explicitly associate Web Service descriptions to
\emph{high-level end-user's intentional requirements}. In our framework,
end-users associate queries to their requirements. Indeed, queries
allow end-users to specify \emph{generic Web Service
descriptions}. For instance, in a neuroscientist community, by looking
for a Web Service which takes as input an image and provides as output a
debiased image, the end-user specifies the kind of Web Service s/he is
interested in without explicitly refering to one specific Web
Service. By doing so, we assume a loosely coupling between high-level
end-user's intentional requirements on one hand and Web Services
descriptions on the other hand: if new Web Service descriptions are
added inside the community Web Service registry, they can be retrieved to
operationalise a high-level end-user's intentional requirement even if
the requirement has been specified before the availability of the Web
Services under consideration; and if Web Service descriptions are
removed from the community Web Service registry, the high-level
end-user's intentional requirements that they satisfied are still valid
and may be operationalised by other available Web Services. Web Services
are dynamically selected when rendering queries associated to
high-level end-user's intentional requirements.

In our approach, we also adopt Web semantic languages and models as a
unified framework to deal with (i) high-level end-user's intentional
requirements, (ii) generic Web Service descriptions and (iii) Web
Service descriptions themselves. With regards to high-level end-user's
intentional requirements, we adapted the map model \cite{Rolland07} to
our concern and gathered its concepts and relationships into an RDFS \cite{rdfs}
ontology dedicated to the representation of intentional processes: the
map ontology  \cite{Corby09}. As a result, intentional processes
annotated with concepts and relationships from this ontology can be
shared and exploited by reasoning on their representations.  We also
consider semantic \emph{Web Service descriptions} specified with the
help of the OWL-S ontology \cite{OWLS}. And finally, \emph{generic Web
Service descriptions} are specified with the help of the W3C standard
query language for RDF \cite{rdf} annotations:
\textsc{sparql} \cite{sparql}. Generic Web Service descriptions
are formalised into graph patterns over Web Services
descriptions. Indeed, our approach relies on three ontologies: The map
ontology we proposed \cite{Corby09}, the OWL-S ontology
\cite{OWLS} and a domain ontology (in our case an ontology
describing medical images and medical image processing dedicated to
the neuroscience domain).

Knowledge capitalisation, management and dissemination inside a
community of members may be supported by a collective memory, that is
to say an explicit, disembodied and persistent representation of the
community knowledge in order to facilitate access, sharing and reuse
\cite{Dieng05}. In semantic collective memories, resources are indexed
by semantic annotations in order to explicit and formalise their
informative content. Information retrieval inside the collective
memory relies on the formal manipulation of these annotations and is
guided by ontologies. In SATIS, we are dealing with annotations about
Web Service descriptions, generic Web Service descriptions and
high-level end-user's intentional requirements. We are exploiting
reasoning and traceability capabilities of semantic Web models and
languages to provide dedicated search, sharing and reuse means to
improve collaboration inside a community of neuroscientists.  Beyond a
way to retrieve Web Services, our approach aims at providing means to
promote mutualisation of high-level end-user's intentional requirements
and cross fertilisation of know-how about how to operationalise image
processing pipelines among the community members. Our proposal may be
compared to case based reasoning approaches in that it provides 
means to identify relevant Web Service descriptions (solutions)
corresponding to new high-level end-user's intentional requirements
(problems) based on Web Service descriptions (solutions) identified
for similar requirements (problems). Indeed high-level end-user's
intentional requirements are considered as problem descriptions and
Web Service descriptions are considered as solutions. generic Web
Service descriptions as well as subgoals and strategies elicited to
specify high-level end-user's intentional requirements are considered as
intermediary knowledge on which to reason to reduce the gap between
high-level end-user's intentional requirements and Web Service
descriptions thus providing solutions to problems that is to say proposing
Web Services to implement an image processing pipeline.

Indeed, we address the
 issue about how to retrieve Web Service descriptions from
high-level end-user's intentional requirements by providing means to
reuse existing knowledge about relevant Web Services to operationalise
high-level end-user's requirements inside the scope of a community of users.

\section{SATIS authoring process}
\label{authoring}

In SATIS, search procedure authoring is supported by a three steps
process: (i) high-level end-user's intentional requirements elicitation,
(ii) requirements and generic Web Service description
formalisation and (iii) fragment definition. During this process, the
map model \cite{Rolland07} helps to capture high-level end-user's
intentional requirements. The map ontology, the domain ontology and
the OWL-S ontology are used to formalise the high-level end-user's
intentional requirements and to specify associated generic Web
Service descriptions. \textsc{RDF} annotations representing high-level
end-user's intentional requirements and \textsc{sparql} queries
formalizing generic Web Service descriptions are then grouped into rules considered as reusable fragments.

\subsection{Elicitation step}

Figure \ref{fig-exemple} shows an example of high-level end-user's
intentional requirement dealing with tissue and lesion
classification. It is specified with the help of the map model
\cite{Rolland07}. According to \cite{Rolland07}, a map is a process
model in which an ordering of \emph{intentions} and
\emph{strategies} has been included.  In our case, we focus on image
processing intentions and image processing strategies. A map is a
labeled directed graph with intentions as nodes and strategies as
edges between intentions. An image processing intention is a goal that
can be achieved by following a strategy. An intention expresses what
is wanted, a state or a result that is expected to be reached
disregarding considerations about who, when and where. There are two
distinct intentions that represent the intentions to start and to stop
the process.  A map consists of a number of sections each of which is
a triple (source intention, target intention, strategy).  A strategy
characterises the flow from the source intention to the target
intention and the way the target intention can be achieved. A map
contains a finite number of paths from its start intention to its stop
intention, each of them prescribing a way to achieve the goal of the
image processing pipeline under consideration. Indeed, it is at
runtime, when an intention is satisfied, that one target intention and
one strategy are chosen (among all the target intention and strategies
available from the current intention), depending on the context of the
process at runtime. 

\begin{figure}[!t]
\centering
\includegraphics[width=3in]{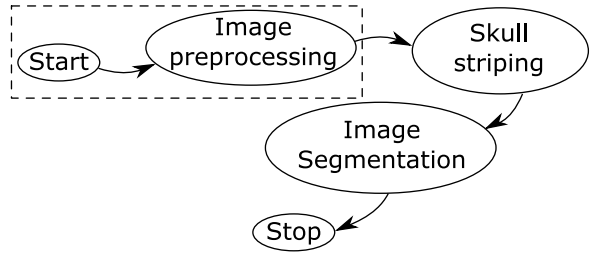}
\caption{Example of high-level end-user's intentional requirement}
\label{fig-exemple}
\end{figure}

In figure \ref{fig-exemple}, we can see 3 main intentions:
\texttt{Image preprocessing}, \texttt{Skull striping} and
\texttt{Image segmentation}. Between the intentions, we discover
strategies. Strategies define the way to pass from an intention to a
next one. There can be many strategies which link up the same
intentions (for instance to indicate which (kind of) algorithm is used
to achieve the target intention). Indeed, in a map, each set which is
made up by a source intention, a strategy and a target intention is a
\emph{section} of the map. An example of section has been highlighted
with a doted line in figure \ref{fig-exemple}. Let's precise that a
map is neither a state diagram, because there is no data structure, no
object, and no assigned value, nor an activity diagram, because there
is always a strong context for each section of the map: its source
intention and its strategy. We can attach more information to this
kind of schema (in order to help the user of the map to choose the
adequate strategy, for example), but this is not the goal of this
paper to fully describe the map model.

The aim of such a modeling is to capture high-level end-user's
intentional requirement in order to turn them into generic Web Service
description to search for available Web Services to implement the
image processing pipeline under consideration. Indeed, high-level
end-user's intentional requirement may need to be further refined to
be transformable into generic Web Service description. For instance,
in the example of figure \ref{fig-exemple}, additional specification
would be useful to understand what kinds of generic Web Service
descriptions are suitable to search for Web Services implementing
image preprocessing. Therefore, each section of a map may be refined
into another map describing more in detail how to reach the target
intention of the section under consideration. Figure
\ref{fig-refinement} shows an example of map refining the section
highlighted in Figure \ref{fig-exemple}. In this map, different ways
(\textit{i.e.} different paths) to achieve the target preprocessing intention
are provided.

\begin{figure}[!t]
\centering
\includegraphics[width=3in]{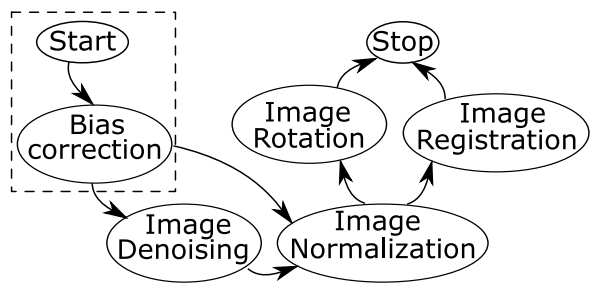}
\caption{Example of requirement refinement}
\label{fig-refinement}
\end{figure}

At the refinement level presented in figure \ref{fig-refinement}, it is
now possible to associate generic Web Service descriptions to map sections in order to specify
how to retrieve Web Service description implementing the section target intention. For instance, a query searching for Web Service descriptions
which have an image as input, and an image qualified as debiased as
well as a bias field as outputs aims at retrieving Web Service
descriptions corresponding to the section highlighted
in figure \ref{fig-refinement}.

\subsection{Formalisation step}
\label{formalisation}
The second step of the authoring process is devoted to the
formalisation of the intentions and strategies elicited during the
previous step, as well as the generic Web Service descriptions
associated to the most refined map sections. Intentions and strategies
are formalised by using verbs, objects and manners from the domain
ontology. Indeed, during the elicitation step, end-users think in terms
of goals and means to reach goals while in the formalisation step,
they try to formalise through domain concepts how to qualify goals and
strategies elicited in the previous step.

To further formalise map elements, we rely on \cite{Prat97,Prat99}
proposal, which has already proved to be useful to formalise goals
\cite{Ralyte01,Guzelian07,Rolland07}. According to
\cite{Prat97,Prat99}, an intention statement is characterised by a
verb and some parameters which play specific roles with respect to the
verb. Among the parameters, there is the object on which the action
described by the verb is processed. We gathered the concepts and
relationships of the map model and this further formalisation into an
RDFS \cite{rdfs} ontology dedicated to the representation of
intentional processes: the map ontology
\cite{Corby09}. \textit{Intention}, \textit{Strategy}, \textit{Verb},
\textit{Object} and \textit{Manner} are examples of concepts provided
in this ontology.

The mappings between the domain ontology and the map ontology are
automatically created when concepts of the domain ontology are
selected to formalise map content. Domain concepts are then considered
as instances of \textit{Verb}, \textit{Object} and
\textit{Manner}. Let us consider again the map depicted in figure
\ref{fig-refinement}. Intention \texttt{Bias correction} is described
by \texttt{Debiasing}, instance of \textit{Verb}, and \texttt{Image},
instance of \textit{Object}. With regards to strategies, up to now we
only consider one strategy between a source and a target
intentions. So far, we did not qualify strategies by binding them to
domain concepts. In the future, we plan to extend our Web Service
annotation model with quality of service (QoS) information and to qualify
map strategies by QoS domain concepts considered as instances of class
\textit{Manner}.

By using the map and the domain ontologies, a common vocabulary is
used by the different members of the community contributing to support
know-how sharing and cross fertilisation.

In SATIS, we assume Web Service descriptions are expressed in
OWL-S. In our current scenarios, we only use the profile and the
grounding of OWL-S as well as the input and output specifications in the
process description. We enrich OWL-S description by considering Web
Service OWL-S description elements (as input and output parameters
for example) as instances of domain concepts. Thanks to this additional instantiation of domain concepts, it makes
it possible to reason on OWL-S description element types to retrieve
for instance subclasses of concepts we are interested in.

Generic Web Service descriptions
 are expressed as \textsc{sparql} queries among the Web Service
descriptions expressed in OWL-S. The following query, where namespace
\texttt{process} refers to the OWL-S process ontology and namespace
\texttt{dom} refers to a domain ontology, is an example of generic Web
Service description to retrieve debiasing Web Service description.

\small
\begin{verbatim}
prefix dom: <http://.../dom-onto#>
prefix process: <http://.../Process.owl#>
select ?service
where 
{         
    ?service process:hasInput ?r1
    filter(?r1 =: dom:Image)
    ?service process:hasOutput ?r2
    filter (?r2 <=: dom:DebiasedImage) 
    ?service process:hasOutput ?r3   
    filter (?r3 <=: dom:BiasField)
}
\end{verbatim}
\normalsize

In this example, we are looking for Web Services which OWL-S
description indicates that the Web Service under consideration
requires a parameter instance of the \texttt{Image} concept from the
domain ontology as input and two parameters, instances of concepts (or
sub-concepts) of \texttt{DebiasedImage} and \texttt{BiasField} as
output.

\subsection{Fragmentation step}

In SATIS, the process consisting in retrieving Web Services
descriptions from high-level end-user's intentional requirements about
image processing pipelines is viewed as a set of loosely coupled
fragments expressed at different levels of granularity. A
fragment is an autonomous and coherent part of a search
process supporting the operationalisation of part of an image
processing pipeline by Web Services. Such a modular view of the
process aiming at retrieving Web Service descriptions from high-level
end-user's intentional requirements favours their adaptation and
extension. Moreover, this view permits to reuse fragments
authored to deal with a specific high-level end-user's image processing
pipeline in the building of other pipelines.

The fragment body captures guidelines that can be
considered as autonomous and reusable. The fragment signature
captures the reuse context in which the fragment can be
applied.

For us, a guideline embodies know-how about how to
achieve an intention in a given situation. We distinguish two types of
guidelines: intentional and operational guidelines. Intentional
guidelines capture high-level end-user's intentional requirements which
have to be refined into more specific requirements. Operational
guidelines capture generic Web Service description.

Map formalisations and \textsc{sparql} queries respectively constitute
the body of intentional and operational reusable fragments. The
fragment signature characterises the fragment content and let the
other members of the community understand in which situation the
fragment may be useful. A fragment signature is specified by a map
section. The target intention of the section indicates the goal of the
reusable fragment and the source intention as well as the strategy
specify the reuse situation in which the fragment is suitable.  The
section highlighted in figure \ref{fig-exemple} is an example of
signature for an intentional fragment which body is the map presented
in figure \ref{fig-refinement}. The section highlighted in figure
\ref{fig-refinement} is an example of signature for an operational
fragment which body is the query presented in section
\ref{formalisation}.

Indeed in SATIS, fragments are implemented by backward chaining
rules, which conclusions represent signatures of fragments and which
premises represent bodies of fragments (either operational or
intentional guidelines). We call
a rule \textit{concrete} or \textit{abstract} depending on whether its
premise encapsulates operational or intentional guidelines.

These rules are implemented as
\textsc{sparql} construct queries. The \textsc{Construct} part is interpreted
as the head of the rule, the consequent that is proved. The \textsc{Where} part
is interpreted as the body, the condition that makes the head
proved. When considered recursively, a set of 
\textsc{sparql} construct queries can be seen as a set of rules processed in
backward chaining.

The following rule, where namespace \texttt{map} refers to
the map ontology, namespace \texttt{process} refers to the OWL-S
ontology and namespace \texttt{dom} refers to a domain ontology, is an
example of concrete rule implementing an operational fragment
 aiming at retrieving debiasing Web Services. 

\small
\begin{verbatim}
<rule rdf:ID="rule-c2">
<rule:value>
prefix dom: <http://.../dom-onto#>
prefix map: <http://.../map-onto#>
prefix process: <http://.../Process.owl#>
construct
{
        _:s map:hasStrategy _:g
        _:g map:hasParameter map:AnyParameter 
        _:s map:hasSource _:o
        _:o map:hasObject map:AnyObject
        _:o map:hasVerb map:AnyVerb
        _:s map:hasTarget _:i
        _:i map:hasObject dom:Image
        _:i map:hasVerb dom:Debiasing
        _:s map:hasResource ?service
}
where 
{         
    ?service process:hasInput ?r1
    filter(?r1 =: dom:Image)
    ?service process:hasOutput ?r2
    filter (?r2 <=: dom:DebiasedImage) 
    ?service process:hasOutput ?r3   
    filter (?r3 <=: dom:BiasField)
}
pragma {cos:server cos:query true}
</rule:value>
</rule>
\end{verbatim}
\normalsize

In the \textsc{Where} part of the rule, we recognise the query
previously presented in this paper. In the \textsc{Construct} part of
the rule, a graph pattern corresponding to the map section to build if
Web Service descriptions are found in the community Web Services
registry is specified. This graph pattern specifies the fact that no
specific strategy and no specific source intention is required to
achieve the target intention (concepts \texttt{map:AnyParameter},
\texttt{map:AnyObject} and \texttt{ map:AnyVerb} are used in the
specification). It also indicates that target intention is formalised
by the object \texttt{dom:Image} and the verb
\texttt{dom:Debiasing}. The retrieved Web Service descriptions are
associated to the newly built map section through the
\texttt{hasResource} property.

Thanks to SATIS three steps authoring process, high-level end-user's
intentional requirements are capitalised inside the community semantic
memory in order to be reused during the rendering process that will be
detailed in the following section.

\section{SATIS rendering process}
\label{rendering}
The \emph{rendering step} is supported by backward chaining among
rules and matching with the Web Service descriptions.  We rely on a
semantic engine for both backward chaining on the SATIS knowledge base
of rules implementing the reusable fragments and matching with the knowledge base of OWL-S Web
Service descriptions. During the
rendering step, high-level end-user's intentional requirements are
dynamically created when needed all along the backward chaining
process, as temporarily subgoals, until Web Service descriptions are
found to match all the sub-goals and therefore the general goal of the
high-level end-user's intentional requirement.  As a result, a community
member looking for solutions to operationalise an image processing
pipeline will take advantage of all the rules and all the Web Service
descriptions stored in the community semantic memory at the time of his/her search. This memory
may evolve over the time and therefore the Web Service descriptions
retrieved by using a rule may vary as well.

Let's clarify that the result is composed of descriptions of candidate
Web Services, and not by Web Services themselves. The invocation of
the selected (among the candidates) Web Services is out of the scope
of this work. When rendering a Web Service descriptions search
process, a set of candidate Web Services (alternatives) is associated
to each goal or subgoal elicited during the specification of the
image processing pipeline. So, the result of the rendering is a
sequence of sets of candidate Web Services. But as the formalism we
choose to model image processing pipeline, the map
model \cite{Rolland07}, allows to specify several way to achieve an
intention, the result of the rendering step may be composed of several
sequences of sets of candidate Web Services.

\section{Improving collaboration among community members}
\label{jungle}

One of the main objectives of SATIS is to support neuroscientists when
looking for Web Services to operationalise their image processing
pipeline. In this section we will first discuss the role of the
different actors involved in the neuroscience community and then
describe the different means we provide to support neuroscientists
tasks.

Three core actors are identified in our framework: the \emph{service
designer}, the \emph{process modeling expert} and the \emph{domain
expert}. In a neuroscientists community, computer scientists play the
roles of \emph{service designer} and \emph{process modeling expert}
while neuroscientists play the role of \emph{domain expert}.

The service designer is in charge of promoting the Web
Services available in the community Web Service registry. Therefore,
when s/he wants to advertise a new kind of Web Service in the
neuroscientists community, in addition to adding the Web Service
description in the community Web Service registry, s/he writes
a generic Web Service description and associates to it high-level
end-user's intentional requirements to promote the services s/he is in
charge from the end-user's point of view (that is to say in a non
computer scientists language, as OWL-S is). The service designer is in charge of
authoring atomic reusable fragments.
 
The process modeling expert is in charge of populating the community
semantic memory with reusable fragments to help domain experts to (i) specify
the image processing pipelines for which they are looking for Web
Services and (ii) search for Web Service descriptions to
operationalise the image processing pipelines they are interested
in. Indeed, s/he provides reusable fragments useful in different image
processing pipelines.  Basic processes, as for instance intensity
corrections, common to several image analysis pipelines, are examples
of such basic fragments. Therefore, s/he may look at the fragments
provided by the service designer with the aim of aggregating some of
them into basic image processing pipelines. For instance, if
 \texttt{Image debiasing}, \texttt{Image denoising}, \texttt{Image
normalisation} and \texttt{Image registration} Web Service
descriptions are provided in the community Web Service registry (and
associated fragments provided in the community semantic memory) at
some point, the process modeling expert may put them together into a
basic \texttt{Image preprocessing} pipeline. S/he may also identify
recurrent needs when supporting domain experts in their authoring task
and therefore provide adequate basic fragments for image processing
pipelines. The process modeling expert may therefore write abstract
rules. If concrete rules about generic Web Service descriptions
corresponding to image processing subgoals are already available,
the process modeling expert only writes the abstract rules. Otherwise,
s/he is also in charge of writing the associated concrete rules.

Finally, the domain expert (or final user) is searching for
Web Service descriptions to operationalise an image processing
pipeline s/he is interested in. Therefore, s/he may first look in the
community semantic memory if some existing rules already deal with the
main intention s/he is interested in. If another member of the
community already authored an image processing pipeline achieving the
same high-level goal, s/he may reuse it as is. The goal under
consideration may also be covered by a larger image processing
pipeline specified through a set of rules already stored in the
community semantic memory and corresponds to one of the subgoals of
the larger pipeline. In this case also, existing rules can be reused
as is and the rendering step to operationalise the image processing
pipeline under consideration performed on the current semantic
community memory content. If no high-level end-user's intentional
requirements are already available, the domain expert specifies the
image processing pipeline under consideration with the help of the
process modeling expert. Indeed, abstract rules
have to be written. 
Then, for each subsection identified in the high-level abstract rule,
the domain expert may search for existing rules supporting their
operationalisation. If it is the case, then s/he can decide to rely on
them and stop the authoring process. Otherwise, s/he
may prefer to provide his/her own way to operationalise the
subgoals. By doing so, the domain expert enriches the semantic
community memory with alternative ways to operationalise already
registered goals. This will result in enriching the
operationalisation means of the image processing pipelines already
formalised into rules stored in the semantic community repository. In
fact, when someone else looking for the subgoals under consideration
will perform a rendering process, if his/her image processing pipeline
relies on the achievement of a target intention for which a new
operationalisation means has been provided, the backward chaining
engine will exploit the rules previously stored in the semantic
community repository as well as the new ones, increasing the number of
ways to find suitable Web Service descriptions. Each time the domain
expert, with the help of the process modeling expert, decides to
provide new ways to operationalise a map section, s/he has to select
the right level of specification of the fragment signature, in
order to allow the reuse of the fragment under construction
outside of the scope of the image processing pipeline under
consideration. 

From a more general point of view, domain and process modeling experts
mainly provide intentional fragments: Domain experts focus on
high-level intentional fragments, close to the image processing
pipelines they want to operationalise. Process modeling experts focus
on low level intentional fragments, that is to say fragments
operationalising basic image processing pipelines. And service
designers mainly focus on providing operational fragments to promote
existing Web Services. But domain and process modeling experts may
also provide operational fragments to specify their requirements in
term of services. And the service designers may also provide
intentional fragments in order to show examples of use of available
Web Services inside the scope of more complex examples of image
processing pipelines. By relying on a rule-based specification to
retrieve Web Service descriptions and by providing distinct and
dedicated modeling techniques to both service designers and service
final-users as well as mapping mechanisms between them, we assist the
bidirectional collaboration between neuroscientists and computer
scientists inside the community.

An important objective of the SATIS project is to provide to domain
experts means to better understand what are the characteristics of the
available services and how to use them in the scope of the image
processing pipeline they are interested in. We support this aim by
several means:

\begin{itemize}

\item Requirements about Web Services are described in terms of
intentions and strategies that is to say a vocabulary familiar to the
domain expert, making the understanding of the a Web Service purpose
easier to understand by domain experts.

\item The SATIS approach relies on a controlled vocabulary (domain
ontology) to qualify Web Services as well as requirements, this way
reducing the diversity in the labeling, especially in Web Services
descriptions elements.

\item We propose to specify required Web Service functionalities in
terms of queries (\textit{i.e.} generic Web Service descriptions) instead
of traditional Web Service descriptions in order to provide an
abstraction level supporting the categorisation of available Web
Services and this way an easier understanding of the content of the
registry by domain experts.

\item In our approach we clearly distinguish an authoring step and a
rendering step:
\begin{itemize}

\item During the authoring step, the focus is on the elicitation of
the search procedure. The domain experts think in terms of intentions
and strategies (and not in terms of services). His/her search procedure
is fully described, eventually with the help of the fragments already
present in the community semantic memory.

\item During the rendering step, it is the system (and not the domain
expert) which tries to find Web Services corresponding to the
requirements specified by the experts (by proving goals and
sub-goals). Indeed, the experts don't need at all to know the content
of the registry. A pertinent subset of it will be extracted by the
system and shown to the experts.  
\end{itemize}

\item And finally, SATIS relies on a rule based approach which doesn't
show to the domain expert the full set of rules exploited by the
backward chaining engine to satisfy the user requirements. When
rendering a search procedure, the domain expert only selects the
intention characterizing his/her image processing pipeline and the
system will search for the rules to use. A set of Web Services
descriptions is given to the domain expert as result. But the
complexity and the number of rules used to get the solution are hidden
to the domain expert.
\end{itemize}

\section{Conclusion}

In this paper, we presented SATIS, a framework to derive Web Service
specifications from end-user's requirements in order to operationalise
business processes in the context of a specific application
domain. More precisely, our framework offers the capability to capture
high-level end-user's requirements in an iterative and incremental way
and to turn them into queries to retrieve Web Services
descriptions. The whole framework relies in reusable and combinable
elements which can be shared out inside the scope of a community of
users. In our approach, we adopt Web semantic languages and models as a
unified framework to deal with (i) high-level end-user's intentional
requirements, (ii) generic Web Service descriptions and (iii) Web
Service descriptions themselves. SATIS aims at
supporting collaboration among the members of a neuroscience
community by contributing to both mutualisation of high-level intentional
specification and cross-fertilisation of know-how about Web Services
 search procedures among the community members.

Future works will first focus on adapting our model to \textsc{corese}
\cite{corese, Corby04}, a semantic Web search engine including a
backward chaining mechanism in order to test our approach on examples
of image processing pipelines. We also plan to develop software tools
in order to automate the main tedious steps, like the transformation
of the map specification into \textsc{sparql} rules and to test our approach in
the context of a neuroscientist community. We also have in mind to
enrich the formalisation step by taking into account additional
information in order, for instance, to derive criteria related to
quality of services. Indeed, we plan to extend our Web Service
annotation model with quality of service (QoS) information and to
qualify map strategies by QoS domain concepts. And we will also
concentrate on providing query patterns to help experts writing
generic Web Service descriptions.

\end{document}